\input{aipcheck}

\documentclass[final]{aipproc}

\layoutstyle{6x9}

\begin{document}

\title{How to calculate the main characteristics of random uncorrelated networks}



\author{Agata Fronczak}{address={Faculty of Physics and Center of Excellence for Complex
Systems Research,\\ Warsaw University of Technology, Koszykowa 75,
PL-00-662 Warsaw, Poland}}

\author{Piotr Fronczak}{address={Faculty of Physics and Center of Excellence for Complex
Systems Research,\\ Warsaw University of Technology, Koszykowa 75,
PL-00-662 Warsaw, Poland}}

\author{Janusz A. Ho\l yst}{address={Faculty of Physics and Center of Excellence for Complex
Systems Research,\\ Warsaw University of Technology, Koszykowa 75,
PL-00-662 Warsaw, Poland}}

\begin{abstract}
We present an analytic formalism describing structural properties
of random uncorrelated networks with arbitrary degree
distributions. The formalism allows to calculate the main network
characteristics like: the position of the phase transition at
which a giant component first forms, the mean component size below
the phase transition, the size of the giant component and the
average path length above the phase transition. We apply the
approach to classical random graphs of Erd\"{o}s and R\'{e}nyi,
single-scale networks with exponential degree distributions and
scale-free networks with arbitrary scaling exponents and
structural cut-offs. In all the cases we obtain a very good
agreement between results of numerical simulations and our
analytical predictions.
\end{abstract}

\maketitle

%%%%%%%%%%%%%%%%%%%%%%%%%%%%%%%%%%%%%%%%%%%%
%% MAINMATTER
%%%%%%%%%%%%%%%%%%%%%%%%%%%%%%%%%%%%%%%%%%%%

\section{Introduction}\label{intro}

\par During the last years, there has been noticed a significant
interest in the field of complex networks and a lot of
interdisciplinary initiatives have been taken aiming at
investigations of these systems \cite{0a,0b,1,14}. It was observed
that despite network diversity, most of real web-like systems
share three prominent characteristics: small-world property (i.e.
small average path length), high clustering and scale-free degree
distribution \cite{32,22}. The observed universal properties let
one understand that networks are not simply a sum of nodes
connected by links. Nowadays, research on networks points mainly
to the so-called emergent properties, that is systems global
features and capabilities which are not specified by network
design and are difficult or impossible to predict from knowledge
of its constituents. As a result, the interest in topological
characterization of real networks gives way to growing interest in
dynamical processes defined on such systems
\cite{BAfluct,BianSOC}. We have already known how networks grow
and how that growth process influences network topology i.e.
pattern of connections \cite{12,VespPRL}. We have also gained a
certain understanding of how network structure mediates different
transport phenomena like: disease (rumor) transmission in social
networks \cite{NewEp} or information processing (virus infection)
in computer networks \cite{PastorEp}. However, there is still a
number of open question like: what are the most efficient and
robust topologies for different web-like systems \cite{BANature}?
How to effectively fight against criminal networks \cite{netwars}?
How to eliminate traffic jams in the Internet and other
transportation systems \cite{TadicPRE}?

\par This paper is devoted to uncorrelated random networks
\cite{NewPRE} (also known as random graphs or configuration model)
which have been repeatedly shown to be very useful in modeling
different phenomena taking place on networks. Although a number of
other network models have been proposed, it was also proved that
in many cases there are no qualitative differences between
phenomena defined on both random graphs and those more
sophisticated models. For example, the main results on percolation
transition \cite{NewPRL1,NewPRL2} and epidemic spreading
\cite{PastorEp,BagunaPRL} that were obtained for random
uncorrelated networks are still valid for correlated and even
non-equilibrium systems. The above certifies that despite its
simplicity, configuration model is not merely mathematical toy.
For the mentioned reasons, the issue of structural properties of
random graphs and their reliable, mathematical description seems
to be very important. It does not only excel in academic
challenges but it can be helpful in further exploration of
networks, including dynamical processes defined on discrete,
web-like topologies.

\par Following the idea, in this paper we proceed with {\it
microscopic} description of the considered networks and ask the
most basic question related to their structural properties i.e.
what is the probability that the shortest distance between two
given nodes $i$ and $j$ is not longer than $x$. Basing on simple,
intuitive arguments we derive the probability and apply it to
calculate the main network characteristics like: the position of
the phase transition at which a giant component first forms
\cite{molloy1,HavPRLpc}, the mean component size below the phase
transition \cite{NewPRE}, the size of the giant component
\cite{molloy2,NewPRL1} and the average path length above the phase
transition \cite{DorMetric}. Although most of the enumerated {\it
macroscopic} properties were previously calculated by means of
other methods, especially thanks to generating function formalism
developed by Newman \cite{NewPRE}, we think that our approach is
conceptually much simpler. Furthermore, since our derivations take
advantage of primary concepts they do not represent a closed
project but can be thought of as a background for further network
investigations, particularly those associated with network
dynamics.

%______________________________________________________________________________ micro
\section{Random uncorrelated networks - microscopic description}\label{micro}

\par Random uncorrelated network with arbitrary degree distribution
$P(k)$ is the simplest network model. In such a network the total
number of vertices $N$ is fixed. Degrees of all vertices are
independent, random integers drawn from a specified distribution
$P(k)$. One can easily create the network following the below
instructions:
\begin{itemize}
\item[i.] Prepare $N$ nodes $i = 1,2,\dots,N$.
\item[ii.] Attach to each node $k_i$ ends of connections taken from
the given distribution $P(k)$.
\item[iii.] Connect at random ends of connections.
\end{itemize}
The above procedure provides maximally random, uncorrelated
networks. From the mathematical point of view the lack of
correlations means that the probability $P(k_i/k_j)$ that an edge
departing from a vertex $j$ of degree $k_j$ arrives at a vertex
$i$ of degree $k_i$, is independent of the initial vertex $j$. The
above translates into the fact that the nearest neighborhood of
each node is the same (in statistical terms). Provided that the
degree distribution in uncorrelated network is $P(k)$ one can show
that the degree distribution of nearest neighbors, that is
equivalent to the probability that an arbitrary link leads to a
node $i$ of degree $k_i$, is given by
\begin{equation}\label{defQ}
P(k_i/k_j)=Q(k_i)=\frac{k_i}{\langle k\rangle}P(k_i),
\end{equation}
and respectively the average nearest neighbor degree equals
\begin{equation}\label{knn}
\langle k_{nn}\rangle=\frac{\langle k^2\rangle}{\langle k\rangle}.
\end{equation}

\par At the moment, before we proceed with our derivations, let us
introduce notation that will be used throughout the paper:
\begin{itemize}
\item[i.] $p^{+}_{ij}(x)$ denotes the probability that there
exists at least one walk of length $x$ between two given nodes $i$
and $j$. If such a walk exists then the shortest path between the
two nodes is $\leq x$.
\item[ii.] $p^{-}_{ij}(x)$ gives the probability that none among
walks of length $x$ occurs between the two nodes, and respectively
\begin{equation}\label{defp-}
p^{-}_{ij}(x)=1-p^{+}_{ij}(x).
\end{equation}
\item[iii.] $p^{*}_{ij}(x)$ describes the probability that the
shortest distance between $i$ and $j$ is equal to $x$. In the
limit of dense networks (above percolation threshold) one can
deduce that if there exists at least one walk of length $x$
between $i$ and $j$ then walks longer than $x$ also exist. It
follows, one can assume that $p_{ij}^{+}(x)$ (see i.) expresses
also the probability that the shortest distance between the two
nodes is not longer than $x$ and
\begin{equation}\label{defp*}
p^{*}_{ij}(x)=p^{+}_{ij}(x)-p^{+}_{ij}(x-1).
\end{equation}
\end{itemize}

\par Now, we try to derive formulas describing the above
probabilities. At the beginning, let us imagine a random walker
that starts at a given node $i$ and assign the walker one-step
memory in order to avoid backward steps. Next, let the walker
perform $x$ steps. It is easy to see that in average the walker
may pass
\begin{equation}\label{Wix}
W(i,x)=k_i(\langle k_{nn}\rangle-1)^{x-1}
\end{equation}
different walks. If, {\it at least} one such a walk ends at the
node $j$ one can say that both vertices $i$ and $j$ are {\it at
most} $x-$th neighbors i.e. the shortest path between the two
nodes is not longer than $x$. It follows that
\begin{equation}\label{p+1}
p^{+}_{ij}(x)=P\left(\bigcup_{r=1}^{W(i,x)} W^{(r)}_{i\rightarrow
j}\right),
\end{equation}
where $W^{(r)}_{i\rightarrow j}$ represents a single walk between
$i$ and $j$, whereas the sum goes over all possible $x-$walks
starting at $i$ (\ref{Wix}).

\par To simplify the last expression (\ref{p+1}) let us assume
mutual independence of all walks $W^{(r)}_{i\rightarrow j}$. Then,
the considered formula may be rewritten in the following form (see
{\it Appendix A})
\begin{equation}\label{p+2}
p^{+}_{ij}(x)\simeq
1-\exp\left[-\sum_{r=1}^{W(i,x)}P\left(W^{(r)}_{i\rightarrow
j}\right)\right].
\end{equation}
To finalize derivation of $p^{+}_{ij}(x)$ note that, due to the
lack of correlations, every walk starting at $i$ has the same
chance to end up in $j$. The probability of such an event is just
equal to the probability that the last connection in the sequence
of edges representing the considered walk is attached to $j$
\begin{equation}\label{PW}
P\left(W^{(r)}_{i\rightarrow j}\right)=\frac{k_j}{\langle k\rangle
N}.
\end{equation}
Inserting the last formula into (\ref{p+2}) and taking advantage
of both (\ref{Wix}) and (\ref{knn}) one gets
\begin{equation}\label{p+3}
p^{+}_{ij}(x)=1-F(x),
\end{equation}
where
\begin{equation}\label{Fx}
F(x)=\exp\left[- \frac{k_{i}k_{j}}{\langle k\rangle
N}\left(\frac{\langle k^2\rangle}{\langle
k\rangle}-1\right)^{x-1}\right].
\end{equation}

\par The last formula for $p^{+}_{ij}(x)$ constitutes the most
important result of the paper. Let us stress that it does not
include any free parameters, therefore one can directly employ it
to calculate all the interesting structural characteristics of the
considered networks. In the following sections we show, how the
formula for percolation threshold in random graphs several times
naturally emerges from our approach. We also calculate, in turn,
the mean component size below the percolation transition, the size
of the giant component and also the average path length above this
transition.

%______________________________________________________________________________ below
\section{Mean component size below percolation threshold}\label{below}

\par To calculate the mean component size below the percolation
threshold one can make use of the introduced probabilities
$p^{+}_{ij}(x)$ and $p^{-}_{ij}(x)$.

\par Taking advantage of (\ref{p+3}), the probability that none
among the walks of length $x$ between the nodes $i$ and $j$ exists
is given by (\ref{defp-})
\begin{equation}\label{pijxC}
p_{ij}^{-}(x)=\exp\left[- \frac{k_{i}k_{j}}{\langle k\rangle
N}\left(\frac{\langle k^2\rangle}{\langle
k\rangle}-1\right)^{x-1}\right],
\end{equation}
and respectively the probability that there is no walk of any
length between these vertices may be written as
\begin{eqnarray}\label{pijm1}
P_{ij}^{-}=\prod_{x=1}^{\infty}p_{ij}^{-}(x)=
\exp\left[-\frac{k_{i}k_{j}}{\langle k\rangle N}
\sum_{y=0}^{\infty}\left(\frac{\langle k^{2}\rangle}{\langle k
\rangle}-1\right)^{y}\right].
\end{eqnarray}
The value of $P_{ij}^{-}$ strongly depends on the common ratio of
the geometric series present in the last equation. When the common
ratio is greater then $1$, i.e. $\langle k^{2}\rangle\geq 2\langle
k\rangle$, random graphs are above the percolation threshold. The
sum of the geometric series in (\ref{pijm1}) tends to infinity and
$P_{ij}^{-}=0$. Below the phase transition, when $\langle
k^{2}\rangle<2\langle k\rangle$, the probability that the nodes
$i$ and $j$ belong to separate clusters is given by
\begin{equation}\label{pijm}
P_{ij}^{-}=\exp\left[-\frac{k_{i}k_{j}}{N}\frac{1}{(2\langle
k\rangle-\langle k^{2}\rangle)}\right],
\end{equation}
and respectively the probability that $i$ and $j$ belong to the
same cluster may be written as
\begin{equation}\label{pijp}
P_{ij}^{+}=1-P_{ij}^{-}=1-\exp\left[-\frac{k_{i}k_{j}}{N}\frac{1}{(2\langle
k\rangle-\langle k^{2}\rangle)}\right].
\end{equation}

\begin{figure}
  \includegraphics[height=.3\textheight]{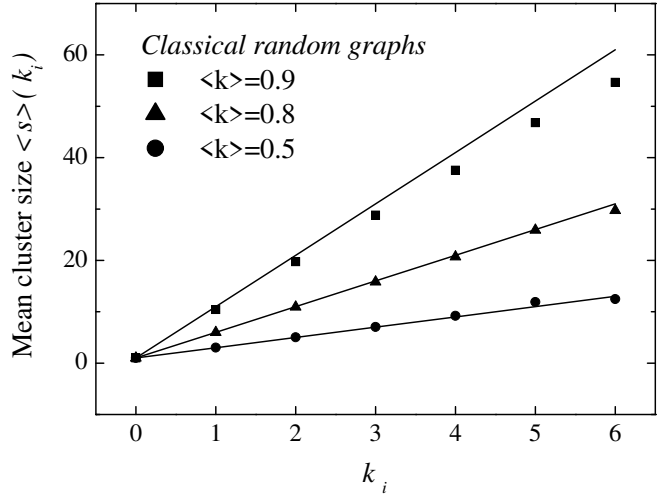}
  \caption{Average size of the component that a node of degree $k$
belongs to. Scatter plots represent numerical data, whereas solid
lines represent the prediction of Eq.
(\ref{skk}).}\label{figERskk}
\end{figure}

\par Now, it is simple to calculate the mean size of the cluster that
the node $i$ belongs to. It is given by
\begin{equation}\label{skk}
\langle s \rangle(k_{i})=1+\sum_{k_{j}}P(k_{j})P_{ij}^{+}\simeq
\frac{\langle k\rangle}{2\langle k\rangle-\langle
k^{2}\rangle}k_{i}+1.
\end{equation}
Note, that the mean size of the component that a node $i$ belongs
to, is a linear function of degree $k_{i}$ of the node (see Fig.
\ref{figERskk}). The last transformation in (\ref{skk}) was
obtained by taking only the first two terms of power series
expansion of the exponential function in (\ref{pijp}). Averaging
the above expression (\ref{skk}) over all nodes in the network one
obtains the known formula \cite{NewPRE} for the mean component
size in random graphs below the phase transition
\begin{equation}\label{sk}
\langle s \rangle=1+\frac{\langle k\rangle^{2}}{2\langle
k\rangle-\langle k^{2}\rangle}.
\end{equation}
As in percolation theory \cite{Stauffer}, the mean cluster size
diverges at
\begin{equation}\label{pc}
\langle k^{2}\rangle=2\langle k\rangle,
\end{equation}
one more time certifying that the expression (\ref{pc}) describes
the position of the percolation threshold in random uncorrelated
networks with arbitrary degree distributions
\cite{molloy1,HavPRLpc}.

%________________________________________________________________________ Giant

\section{Size of the giant component}

\par When $\langle k^{2}\rangle>2\langle k\rangle$ the giant component
(GC) is present in the graph. Its relative size $S$, i.e. the
probability that a node belongs to GC, is an important quantity
and is often identified as the order parameter of the percolation
transition. In this section, we demonstrate how to calculate the
size of the giant component $S$ in uncorrelated networks with
arbitrary degree distributions. The underlying concept of our
derivations is closely related to the method of calculating $S$ in
Cayley tree and originates from Flory (1941)
\cite{Flory,Stauffer}.

\par At the beginning, we deal with classical random graphs of
Erd\"{o}s and R\'{e}nyi, then we generalize our calculations for
the case of random graphs with arbitrary degree distributions and
we show that our derivations are consistent with the formalism
based on generating functions that was introduced by Newman {\it
et al.} \cite{NewPRE}.

%________________________________________________________________________ GC ER
\subsection{Classical random graphs of Erd\"{o}s and R\'{e}nyi (ER)}

\par In general, one defines the classical random graph as N
labeled nodes and every pair of the nodes being connected with
probability $p$ \cite{Bollobas}. The probability that an arbitrary
node $i$ belongs to the giant component is equivalent to the
probability that at least one of its $N-1$ possible links connects
it to GC. In order to describe the above equivalence by means of
mathematical expression, let us assume that $A_{\{i,j\}}$
represents an event: the connection $\{i,j\}$ (if exists!) leads
to the giant component. The notation $A_{\{i,j\}}$ refers to each
of $N(N-1)/2$ possible links, not only to those existing! Now, the
size of the giant component is given by
\begin{equation}\label{ERS1}
S=P\left(\bigcup_{j=1}^{N-1}A_{\{i,j\}}\right),
\end{equation}
and due to the mutual independence of different links, the last
formula may be rewritten as (see {\it Appendix A})
\begin{equation}\label{ERS2}
S=1-\exp\left[-\sum_{j=1}^{N-1}P(A_{\{i,j\}})\right].
\end{equation}
Next, to further simplify the expression for $S$ note that the
mentioned mutual independence implies that every link has the same
probability to belong to the infinite cluster i.e.
\begin{equation}\label{defR}
\forall_{\{i,j\}}\;\; P(A_{\{i,j\}})=R.
\end{equation}
Inserting (\ref{defR}) into (\ref{ERS2}) one gets
\begin{equation}\label{Ser}
S=1-\exp[-(N-1)R].
\end{equation}

\par To finalize derivation of $S$ one has to find $R$. To do so, let
us recall that $R$ describes the probability that an arbitrary
node $i$ is connected to the giant component through a fixed link
$\{i,j\}$, where $j$ is another arbitrary node. Now, if $j$
belongs to GC it means that at least one of its connections also
leads to the giant component. Given that every node may have $N-1$
links the formula for $R$ may be expressed as a product of the
probability of a link $\{i,j\}$ and the probability that at least
one of $N-1$ possible links emanating from $j$ connects the
considered node to the giant component. Summarizing the above
considerations one obtains a self-consistency equation for $R$
\begin{eqnarray}\label{Rer}
R=P(A_{\{i,j\}})=pP\left(\bigcup_{m=1}^{N-1}A_{\{j,m\}}\right)\\\nonumber=p(1-\exp[-(N-1)R]).\;
\end{eqnarray}

\begin{figure}
  \includegraphics[height=.3\textheight]{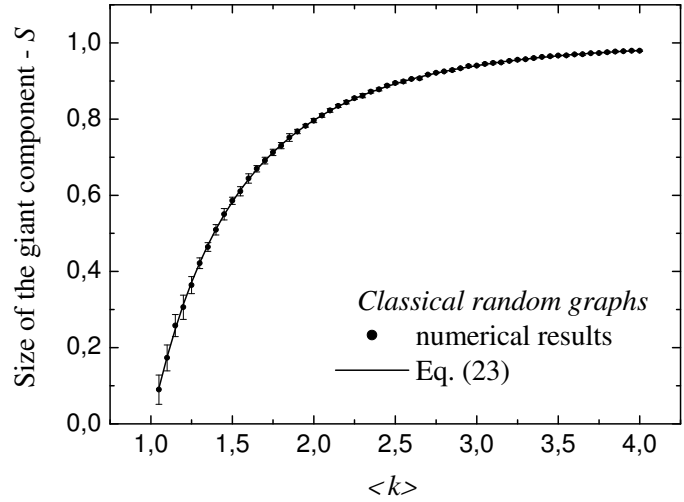}
\caption{Size of the giant component $S$ versus the average degree
$\langle k\rangle$ in classical random graphs of size $N=10000$.
The scatter plot represents numerical data whereas the solid line
gives the solution of the Eq. (\ref{Serfinal})}. \label{figERS}
\end{figure}

\par
\begin{figure}
  \includegraphics[height=.3\textheight]{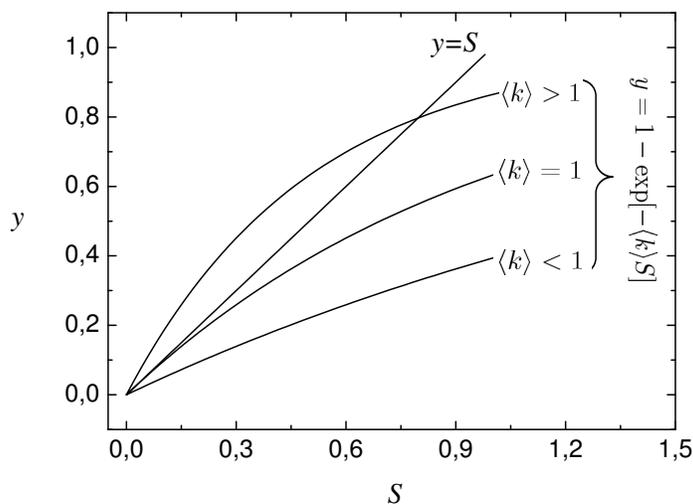}
\caption{Graphical solution of the Eq. (\ref{Serfinal}).}
\label{figERpc}
\end{figure}

 Finally, comparing both relations (\ref{Ser}) and (\ref{Rer}) it
is easy to see that $R=pS$ and the expression for the giant
component in classical random graphs may be rewritten in the
following form \cite{molloy2}
\begin{equation}\label{Serfinal}
S=1-\exp[-\langle k\rangle S],
\end{equation}
where $\langle k\rangle=pN$ (see Fig. \ref{figERS}). At the moment
let us point a certain interesting property of the last equation,
that makes the equation very intuitive example of percolation
transition (at least for those acquainted with Ising model). Below
the percolation threshold (i.e. for $\langle k\rangle<1$
\footnote{Degree distribution in classical random graphs is
Poissonian, therefore the expression for percolation threshold
(\ref{pc}) simplifies to $\langle k \rangle=1$.}) the identity has
only one solution $S=0$ (see Fig. \ref{figERpc}). Above the
threshold another solution $S\neq 0$ appears signifying transition
of the system to the ordered phase.

%________________________________________________________________________ GC Arbitrary
\subsection{Giant component in random graphs with arbitrary degree distributions}
\par In this section, taking advantage of the intuition gained from the
analysis of the giant component in classical random graphs, we
develop a more general approach allowing to calculate the size of
the giant cluster in random networks with arbitrary degree
distributions $P(k)$.

\par In the case of classical random graphs all nodes have been
considered equivalent. It is not acceptable in the case of random
graphs with a given degree sequence $P(k)$, where every node has a
fixed number of connections. In order to meet the requirements
imposed by $P(k)$, we have to slightly rearrange the previous
meaning of the probability $R$ (\ref{defR}). Now, let us assume
that $R$ describes the probability that an arbitrary but existing
(!) connection belongs to the giant component. It is also useful
to think of $R$ as the probability that following arbitrary
direction of a randomly chosen edge one arrives at the giant
component. In fact, we know that following an arbitrary edge we
arrive at a vertex of degree $k$. The probability that the node,
we have just arrived at, is connected to GC equals
$1-(1-R)^{k-1}$. The last relation simply expresses the
probability that at least one of $k-1$ edges emanating from $i$
and other than the edge we arrived along connects $i$ to the giant
component \footnote{Here, we do not take advantage of the Lemma
\ref{tw1} because it works well only in the limit of large number
of the contributing events $n\gg 1$. In the case of small $n$ the
error $q$ of the Lemma \ref{tw1} can not be neglected.}.
Summarizing the above considerations makes simple to write the
self-consistency condition for $R$ (compare with Eq. (\ref{Rer}))
\begin{equation}\label{Rrg}
R=\sum_{k}\left(1-(1-R)^{k-1}\right)Q(k),
\end{equation}
where $Q(k)$ is given by (\ref{defQ}). Then, knowing $R$ it is
easy to calculate the relative size $S$ of the giant component
that is equivalent to the probability that at least one of $k$
links attached to an arbitrary node connects the node to GC
\begin{equation}\label{Srg}
S=\sum_{k}\left(1-(1-R)^{k}\right)P(k).
\end{equation}

\par To make derivations of this section more concrete, we should
immediately introduce some examples of specific networks. Since
however, one can show that both formulas (\ref{Srg}) and
(\ref{Rrg}) are completely equivalent to equations derived by
other authors (see {\it Appendix B}), we just refer the reader to
analyze examples presented in those related papers
\cite{NewPRE,NewSIAM}.

%________________________________________________________________________ APL
\section{Average path length in random uncorrelated networks}

\par We turn now to the quantitative description of the small-world
effect in random uncorrelated networks with arbitrary degree
distributions $P(k)$. To our knowledge, the below derivations are
the simplest and the most accurate among those previously reported
\cite{NewPRE,MotterPRE,DorMetric}, therefore we illustrate them
with a larger number of examples.

\par Taking advantage of (\ref{p+3}), we are in a position to
calculate $p_{ij}^*(x)$ (\ref{defp*}) i.e. probability
distribution for the shortest distance between any two nodes $i$
and $j$
\begin{equation}\label{p*1}
p_{ij}^{*}(x)=F(x-1)-F(x),
\end{equation}
where $F(x)$ is given by (\ref{Fx}). Although, due to the
condition $p_{ij}^{*}(x)\geq 0$ the last expression is only
correct above the percolation threshold, the formula (\ref{p*1})
is very important and, given that the giant component contains
almost all vertices, it allows to find a number of structural
properties of the considered networks. For example, averaging
(\ref{p*1}) over all pairs of nodes one obtains the intervertex
distance distribution $p(x)=\langle\langle
p_{ij}^{*}(x)\rangle_{i} \rangle_{j}$. It is also possible to
calculate $z_{x}$ i.e. the mean number of vertices a certain
distance $x$ away from a randomly chosen node $i$. The quantity
can be obtained as $z_{x}=N\int p_{ij}^{*}(x)P(k_{j})dk_{j}$. At
the moment, let us note that taking only the first two terms of
power series expansion of both exponential functions $F$ in
(\ref{p*1}) one gets the relationship
$z_{x}=z_{1}(z_{2}/z_{1})^{x-1}=\langle k \rangle(\langle
k_{nn}\rangle-1)^{x-1}$ (compare it with (\ref{Wix})) that was
obtained by other authors given the assumption of a tree-like
structure of random graphs. Finally, the average path length (APL)
between any two nodes $i$ and $j$ of degrees $k_i$ and $k_j$ is
simply the expectation value of the distribution (\ref{p*1})
\begin{equation}\label{lij1}
l_{ij}(k_{i},k_{j})=\sum_{x=1}^{\infty}xp_{ij}^{*}(x)=\sum_{x=0}^{\infty}F(x).
\end{equation}
The Poisson summation formula allows one to simplify the above sum
(see {\it Appendix C})
\begin{equation}\label{lij}
l_{ij}(k_{i},k_{j})=\frac{-\ln(k_{i}k_{j})+\ln(\langle
k^{2}\rangle-\langle k\rangle)+\ln N-\gamma}{\ln(\langle
k^{2}\rangle/\langle k\rangle-1)}+\frac{1}{2},
\end{equation}
where $\gamma\simeq 0.5772$ is the Euler's constant. The average
intervertex distance for the whole network depends on a specified
degree distribution $P(k)$
\begin{equation}\label{lRG}
l=\frac{-2\langle\ln k\rangle+\ln(\langle k^{2}\rangle-\langle
k\rangle)+\ln N-\gamma}{\ln(\langle k^{2}\rangle/\langle
k\rangle-1)}+\frac{1}{2}.
\end{equation}
As one could expect, the both formulas (\ref{lij}) and (\ref{lRG})
diverge when the considered networks approach percolation
threshold $\langle k^2\rangle=2\langle k\rangle$ (\ref{pc}).

\par To test the formula (\ref{lRG}) we have performed numerical
simulations of the well-known networks: classical random graphs
proposed by Erd\"os and R\'enyi (ER), single-scale networks with
exponential degree distribution (EXP) and scale-free networks
(SF).

\begin{figure}
  \includegraphics[height=.3\textheight]{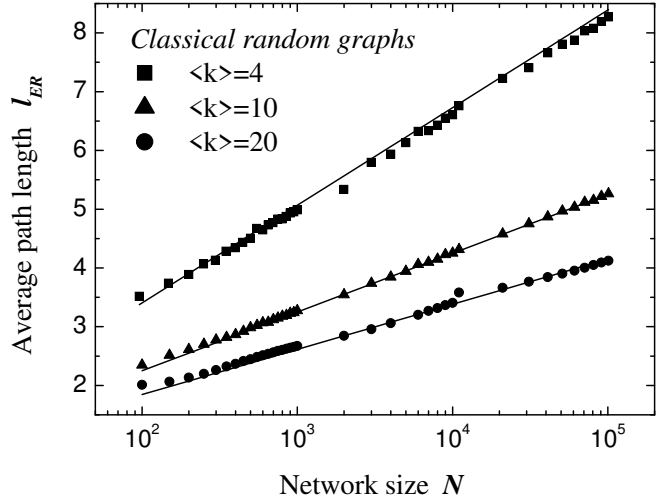}
\caption{Average path length $l_{ER}$ versus network size $N$ in
$ER$ classical random graphs with $\langle k \rangle =pN=4,10,20$.
The solid curves represent numerical prediction of Eq.
(\ref{lER}).} \label{figer}
\end{figure}

\par {\it Classical random graphs (ER)}. For these networks the degree
distribution is Poissonian
\begin{equation}\label{Poisson}
P(k)=\frac{e^{-\langle k\rangle}\langle k\rangle^{k}}{k!}.
\end{equation}
However, since $\langle\ln k\rangle$ cannot be calculated
analytically for Poisson distribution thus the $APL$ may not be
directly obtained from (\ref{lRG}). To overcome this problem we
take advantage of the mean field approximation and assume that all
vertices within a graph possess the same degree
$\forall_{i}\:k_{i}=\langle k\rangle$. It implies that the $APL$
between two arbitrary nodes $i$ and $j$ (\ref{lij}) also describes
the average intervertex distance characterizing the whole network
\begin{equation}\label{lER}
l_{ER}=\frac{\ln N - \gamma}{\ln(pN)}+\frac{1}{2}.
\end{equation}

\par Until now only a rough estimation of the quantity has been
known. One has expected that the average shortest path length of
the whole ER graph scales with the number of nodes in the same way
as the network diameter. We remind that the diameter $d$ of a
graph is defined as the maximal distance between any pair of
vertices and  $d_{ER}=\ln N/\ln(pN)$ \cite{1}. Figure \ref{figer}
presents the prediction of the equation (\ref{lER}) in comparison
with the numerically calculated APL in classical random graphs.

\begin{figure}
  \includegraphics[height=.3\textheight]{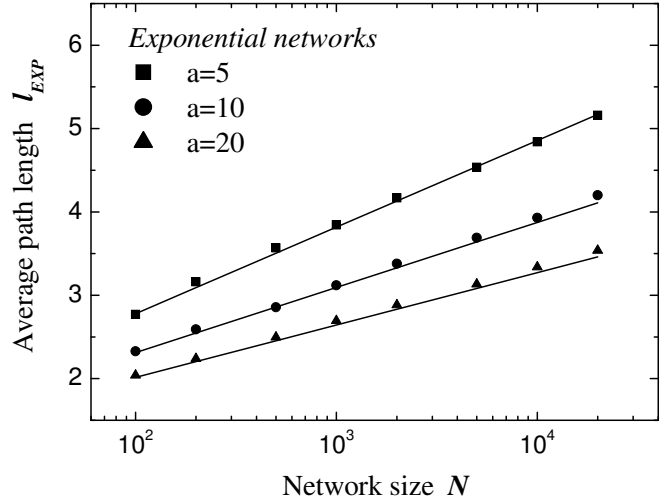}
\caption{Characteristic path length $l_{EXP}$ versus network size
$N$ in exponential networks. Solid lines represent Eq.
(\ref{lEXP}).} \label{figexpo}
\end{figure}

\par {\it Exponential networks (EXP)}. Now, let us suppose that the
degree distribution is exponential
\begin{equation}\label{EXPPk}
P(k)=\frac{e^{(1-k)/a}}{a},
\end{equation}
where $k=1,2,\dots$ \footnote{We start to enumerate degrees from
$k=1$ instead of $k=0$ in order to prevent construction of very
sparse networks with a large number of separated nodes.}. Applying
the distribution to Eq. (\ref{lRG}) immediately provides the
formula for the average path length in the considered single-scale
networks
\begin{equation}\label{lEXP}
l_{EXP}=\frac{\ln
N+\ln(2a^2+a)-2e^{1/a}\Gamma(0,1/a)-\gamma}{\ln(2a^2+a)-\ln(a+1)}+\frac{1}{2},
\end{equation}
where $\Gamma(0,1/a)$ is an incomplete gamma function. Figure
\ref{figexpo} shows that the obtained formula perfectly fits
numerical results obtained for different values of the parameter
$a$.

\par {\it Scale-free networks (SF)}. As mention at the beginning of
the paper degree distributions are scale-free in most of real
systems
\begin{equation}\label{SFPk}
P(k)=\frac{(\alpha-1)m^{(\alpha-1)}}{k^{\alpha}},
\end{equation}
where $k=m,m+1,\dots,k_{max}$. It was found that the most generic
mechanism driving real networks into scale-free structures is the
linear preferential attachment. The simplest model that
incorporates the rule of preferential attachment was introduced by
Barab\'{a}si and Albert \cite{22}. Other interesting mechanisms
leading to scale-free networks were proposed by Goh et al.
\cite{gohPRL2001} and Caldarelli et al. \cite{calPRL2002}.
Unfortunately, the mentioned mechanisms leading to scale-free
network topologies incorporate structural correlations, making our
approach useless. The simple procedure of generating random
uncorrelated networks that was described at the beginning of this
paper also fails when applied to fat-tailed degree distributions
with diverging second moment $\langle k^2\rangle$. In particular,
the procedure may not be applied to generate uncorrelated
scale-free networks (\ref{SFPk}) with the scaling exponent
$2<\alpha<3$. In order to avoid the inconvenience connected with
those scale-free distributions we apply the so-called structural
cutoffs by imposing the largest degree to scale as
$k_{max}=\sqrt{\langle k\rangle N}$
\cite{BurdaPRE,BogunaEPJB,CatanzaroPRE}, instead of unbounded
cutoff predicted by extreme value theory $k_{max}\sim
N^{1/(\alpha-1)}$.

\par
\begin{figure}
  \includegraphics[height=.9\textheight]{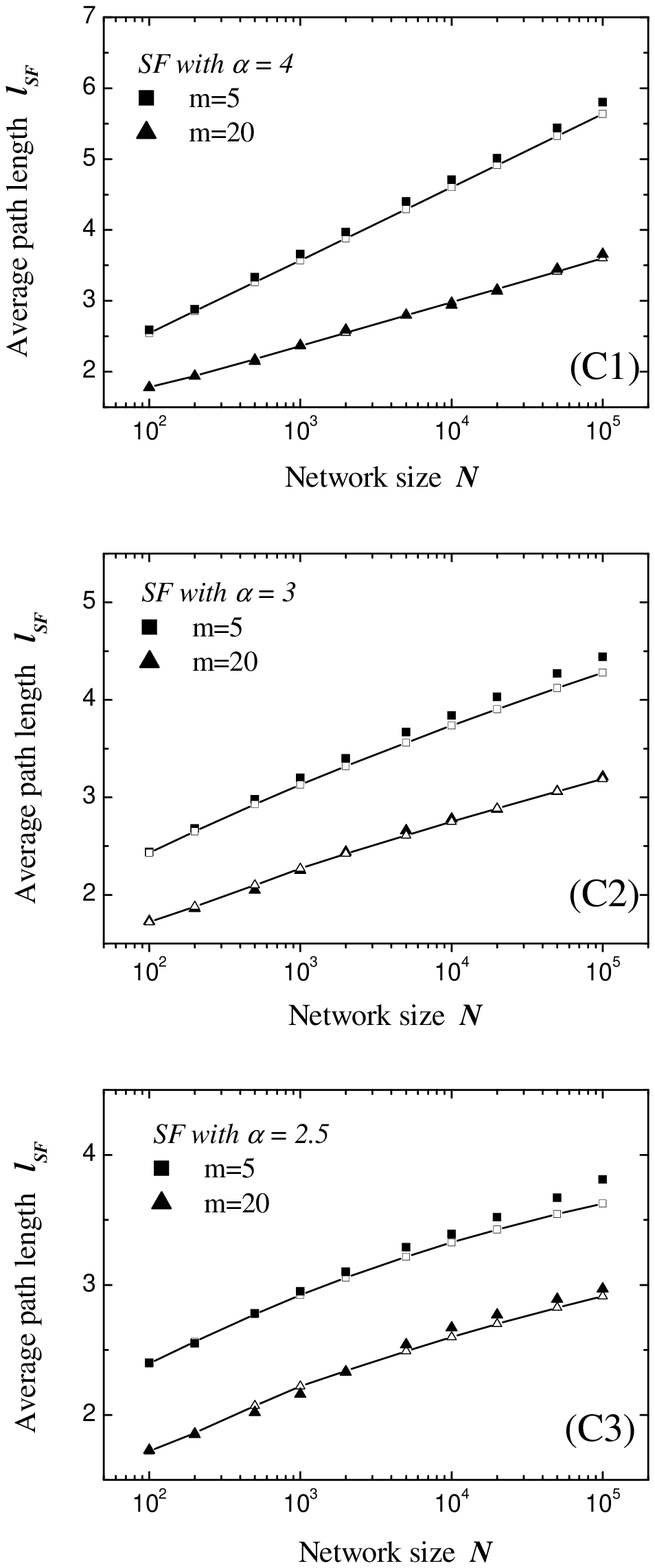}
\caption{Average path length versus number of nodes $N$ in
scale-free networks with $\alpha=4$ (C1), $\alpha=3$ (C2) and
$\alpha=2.5$ (C3). The scatter data represent numerical
calculations. Solid curves with open symbols express analytical
predictions of Eq. (\ref{lER}).} \label{figsf}
\end{figure}

 We have found that depending on the value of the exponent $\alpha$ one can
distinguish three scaling regions for the average path length
(\ref{lRG}) in scale-free networks (see Fig. \ref{figsf}). In the
limit of large systems $N\rightarrow\infty$, $APL$ scales
according to relations \footnote{The exact formulas for APL in
those regions are very complicated therefore we have decided to
omit them.}
\begin{itemize}
\item for $\alpha>3$
\begin{equation}\label{APLa4N}
l^{\alpha>3}\sim \ln N,
\end{equation}
\item for $\alpha=3$
\begin{equation}\label{APLa3N}
l^{\alpha=3}\sim \frac{\ln N}{\ln\ln N},
\end{equation}
\item for $2<\alpha<3$
\begin{equation}\label{APLa25N}
l^{\alpha<3} = const.
\end{equation}
\end{itemize}

\par Note that although the results for $\alpha\geq 3$ are
consistent with estimations obtained by other authors
\cite{DorMetric,HavPRLultra}, the case of $2<\alpha<3$ is
different. In opposite to previous estimations signaling the
double logarithmic dependence $l^{\alpha<3}\sim\ln\ln N$, our
calculations for the same range of $\alpha$ predict that there is
a saturation effect for the mean path length in large networks.
Unfortunately, at the moment it is impossible to assess what is
the correct estimation (to perform reliable tests very large
networks, even with $N>10^{10}$, should be analyzed). On the other
hand, it is truth that looking at Figure \ref{figsf} one can
observe two effects: i. the difference between results of
numerical calculations and our analytical predictions continuously
grows with $N$,  ii. denser networks are better described by our
approach. The first effect may result form the fact that our
mean-field derivations work worse for networks with degree
distributions characterized by large fluctuations (note that the
second moment of scale-free distribution described by the exponent
$\alpha\leq 3$ increases with $N$), whereas the second effect may
be related to the fact that our approach (in particular Eq.
(\ref{p*1})) works better for networks further above the
percolation threshold

\section{Conclusions}

\par To conclude, in this paper we have presented theoretical approach
for metric properties of uncorrelated random networks with
arbitrary degree distributions. We have derived a formula for
probability $p^{+}_{ij}(x)$ (\ref{p+3}) that there exists at least
one walk of length $x$ between two arbitrary nodes $i$ and $j$. We
have shown that given $p^{+}_{ij}(x)$ one can find a number of
structural characteristics of the studied networks. In particular,
we have applied our approach to calculate the mean component size
below the percolation transition, the size of the giant component
and the average path length above the phase transition. We have
successfully applied our formalism to such diverse networks like
classical random graphs of Erd\"{o}s and R\'{e}nyi, single-scale
networks with exponential degree distributions and uncorrelated
scale-free networks with structural cut-offs. In all the studied
cases we noticed a very good agreement between our theoretical
predictions and results of numerical investigation.

\section{Appendices}

\subsection{Appendix A}

\newtheorem{tw}{Lemma}
\begin{tw}\label{tw1}
If $A_{1},A_{2},\dots,A_{n}$ are mutually independent events and
their probabilities fulfill relations $\forall_{i} P(A_{i})\leq
\varepsilon$ then
\begin{equation}\label{pf}
P(\bigcup_{i=1}^{n}A_{i})=1-\exp(-\sum_{i=1}^{n}P(A_{i}))-Q,
\end{equation}
where $0\leq Q<\sum_{j=0}^{n+1}
(n\varepsilon)^{j}/j!-(1+\varepsilon)^{n}$.
\end{tw}
{\bf Proof.} Using the method of inclusion and exclusion
\cite{Feller} we get
\begin{eqnarray}\label{p1}
P(\bigcup_{i=1}^{n}A_{i})=\sum_{j=1}^{n}(-1)^{j+1}S(j),
\end{eqnarray}
with
\begin{eqnarray}\label{p2}
S(j)=\sum_{1 \leq i_{1} < i_{2} < \dots < i_{j} \leq n}^{n}
P(A_{i_{1}})P(A_{i_{2}})\dots P(A_{i_{j}}) \nonumber \\
=\frac{1}{j!}\left ( \sum_{i=1}^{n} P(A_{i}) \right )^{j}-Q_{j},
\end{eqnarray}
where $0\leq Q_{j}\leq \left( n^{j}/j! - \left( ^{n}_{j}
\right)\right)\varepsilon^{j}$. The term in bracket represents the
total number of redundant components occurring in the last line of
(\ref{p2}). Neglecting $Q_{j}$ it is easy to see that $(1-P(\cup
A_{i}))$ corresponds to the first $(n+1)$ terms in the MacLaurin
expansion of $\exp(-\sum P(A_{i}))$. The effect of higher-order
terms in this expansion is smaller than
$R<(n\varepsilon)^{n+1}/(n+1)!$. It follows that the total error
of (\ref{pf}) may be estimated as $Q<\sum_{j=1}^{n}Q_{j}+R$. This
completes the proof.

Let us notice that the terms $Q_{j}$ in (\ref{p2}) disappear when
one approximates multiple sums $\sum_{1 \leq i_{1} < i_{2} < \dots
< i_{j} \leq n}^{n}$ by corresponding multiple integrals. For
$\varepsilon = A/n\ll 1$ the error of the above assessment is less
then $A^{2}\exp(A)/n$ and may be dropped in the limit
$n\rightarrow \infty$.

\subsection{Appendix B}

It is easy to show that our formulas (\ref{Srg}) and (\ref{Rrg})
are completely equivalent to equations derived by Newman et al.
\cite{NewPRE} by means of generating functions
\begin{equation}\label{new1}
S=1-G_{0}(v),
\end{equation}
where $v$ is the solution of equation given below
\begin{equation}\label{new2}
v=G_{1}(v).
\end{equation}
We recall that $G_{0}(x)$ is the generating function for the
degree distribution
\begin{equation}\label{G0}
G_{0}(x)=\sum_{k}P(k)x^{k},
\end{equation}
whereas $G_{1}(x)$ is related to $Q(k)$ (\ref{defQ})
\begin{equation}\label{G1}
G_{1}(x)=\frac{1}{\langle k\rangle}\frac{dG_{0}(x)}{dx}
=\sum_{k}Q(k)x^{k-1}.
\end{equation}

Let us start with Eq. (\ref{Rrg}) that may be transformed in the
following way
\begin{equation}
R=\sum_{k} Q(k) - \sum_{k}Q(k)(1-R)^{k-1} =1-G_{1}(1-R).
\end{equation}
Note, that the last formula directly corresponds to Eq.
(\ref{new2}) with
\begin{equation}\label{defv}
v=1-R.
\end{equation}
Taking into account the last identity one can show that the
expression (\ref{Srg}) may be transformed into Eq. (\ref{new1}) in
a similar way. Now, it is clear that the unknown parameter $v$ in
both Eqs. (\ref{new1}) and (\ref{new2}) has the following meaning
- it describes the probability that an arbitrary edge in a random
graph does not belong to the giant component.

\subsection{Appendix C}

The Poisson summation formula states
%\begin{eqnarray*}
%\sum_{x=0}^{\infty}F(x)=\frac{1}{2}F(0)+\int_{0}^{\infty}F(x)dx
%+2\sum_{n=1}^{\infty}\left(\int_{0}^{\infty}F(x)\cos(2n\pi
%x)dx\right).
%\end{eqnarray*}
\begin{eqnarray}\label{poissonform}
\sum_{x=0}^{\infty}F(x)=\frac{1}{2}F(0)+\;\;\;\;\;\;\;\;\;\;\;\;\;\;
\;\;\;\;\;\;\;\;\;\;\;\;\;\;
\;\;\;\;\;\;\;\;\;\;\;\;\;\;\;\;\;\;\;\;\;\;\;\;\;\;
\\\;\;\;\;\;\;\;\;\;\;\;\;\;\;\;
\int_{0}^{\infty}F(x)dx
+2\sum_{n=1}^{\infty}\left(\int_{0}^{\infty}F(x)\cos(2n\pi
x)dx\right)\nonumber.
\end{eqnarray}
Applying the formula to (\ref{lij})
\begin{eqnarray}
l_{ij}(k_{i},k_{j})=\sum_{x=0}^{\infty}\exp\left[-\frac{k_{i}k_{j}}{(\langle
k^{2} \rangle-\langle k\rangle)N}\left(\frac{\langle
k^{2}\rangle}{\langle k\rangle}-1\right )^{x\;}\right]
\end{eqnarray}
one realizes that in most of cases
\begin{eqnarray}
\frac{k_{i}k_{j}}{(\langle k^{2}\rangle-\langle k\rangle) N}\simeq
0
\end{eqnarray}
that gives $F(0)=1$. One can also find that
\begin{eqnarray}
\int_{0}^{\infty}F(x)dx=-Ei\left(-\frac{k_{i}k_{j}}{(\langle
k^{2}\rangle-\langle k\rangle)N}\right)/\ln\left(\frac{\langle
k^{2}\rangle}{\langle k\rangle}-1\right),
\end{eqnarray}
where $Ei(y)$ is the exponential integral function that for
negative arguments is given by $Ei(-y)=\gamma+\ln y$ \cite{Ryzyk},
where $\gamma\simeq 0.5772$ is the Euler's constant. Finally, it
is easy to see that owing to the generalized mean value theorem
every integral in the last term of the summation formula
(\ref{poissonform}) is equal to zero. It follows that the equation
for $l_{ij}(k_{i},k_{j})$ is given by (\ref{lij}).

\begin{theacknowledgments}
This work has been partially supported by European Commission
Project CREEN FP6-2003-NEST-Path-012864 and by the COST Programme
P10 {\it Physics of Risk}. A.F. and J.A.H. acknowledge the State
Committee for Scientific Research in Poland for financial support
under grant No. 1P03B04727.
\end{theacknowledgments}

%\begin{references}

\end{document}